\documentclass{article}
\usepackage{spconf}

\usepackage[linesnumbered,ruled,vlined]{algorithm2e}
\usepackage{multirow}
\usepackage{multicol}
\usepackage{xcolor}

\usepackage{graphicx}
\usepackage{amsmath}

\usepackage{amssymb}
\usepackage{booktabs}
\usepackage{appendix}
\usepackage{float}
\usepackage[labelformat=simple]{subcaption}

\usepackage{mathtools}
\usepackage{url}
\usepackage{tabularx}
\usepackage{footmisc}
\usepackage{array}
\usepackage[flushleft]{threeparttable}
\usepackage{hyperref}
\usepackage{cleveref}

\usepackage{balance}

\copyrightnotice{\copyright\ IEEE 2023}
\toappear{To appear in {\it Proc.\ ICASSP2023, June 04-10, 2023, Rhodes Island, Greece}}

\crefname{section}{Sec.}{Secs.}
\crefname{table}{Tab.}{Tabs.}
\crefname{figure}{Fig.}{Figs.}

\newcommand{\eg}{\emph{e.g.,}~}

\newcommand{\ie}{\emph{i.e.,}~}
\newcommand{\wrt}{\emph{w.r.t.}~}

\newcommand{\ModelName}{THA\xspace}%
\newcommand{\TTHImage}{TH\xspace}%

\def\clipF{{{\color{black} CLIP-\textit{flickr}}}}
\def\clipC{{{\color{black} CLIP-\textit{coco}}}}

\title{Towards Making a Trojan-horse Attack on Text-to-Image Retrieval}
\name{Fan Hu~~~~~~Aozhu Chen~~~~~~Xirong Li\sthanks{Corresponding author: Xirong Li (xirong@ruc.edu.cn)}} 
 
\address{AIMC Group, MoE Key Lab of DEKE, Renmin University of China 
}

\begin{document}
%
\maketitle

\begin{abstract}

While deep learning based image retrieval is reported to be vulnerable to adversarial attacks, existing works are mainly on image-to-image retrieval with their attacks performed at the front end via query modification. By contrast,  we present in this paper the first study about a threat that occurs at the back end of a text-to-image retrieval (T2IR) system. Our study is motivated by the fact that the image collection indexed by the system will be regularly updated due to the arrival of new images from various sources such as web crawlers and advertisers. With malicious images indexed, it is possible for an attacker to indirectly interfere with the retrieval process, letting users see certain images that are completely irrelevant w.r.t. their queries. We put this thought into practice by proposing a novel Trojan-horse attack (THA). In particular, we construct a set of Trojan-horse  images by first embedding word-specific adversarial information into a QR code and then putting the code on benign advertising images. A proof-of-concept evaluation,  conducted on two popular T2IR datasets (Flickr30k and MS-COCO), shows the effectiveness of the proposed THA in a white-box mode.

\end{abstract}

\section{Introduction} \label{sec:intro}


\begin{figure}[hb!]
  \begin{subfigure}{\columnwidth}
    \centering
    \includegraphics[width=\columnwidth]{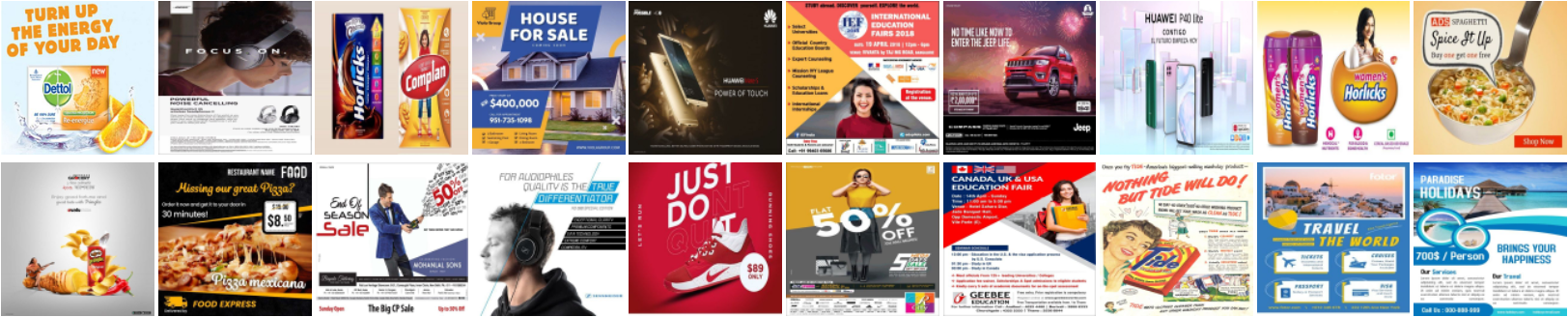} 
    \caption{Twenty benign images, randomly gathered from the Internet}
    \label{fig:benign}
  \end{subfigure}
   
   \begin{subfigure}{\columnwidth}
    \centering
    \includegraphics[width=\columnwidth]{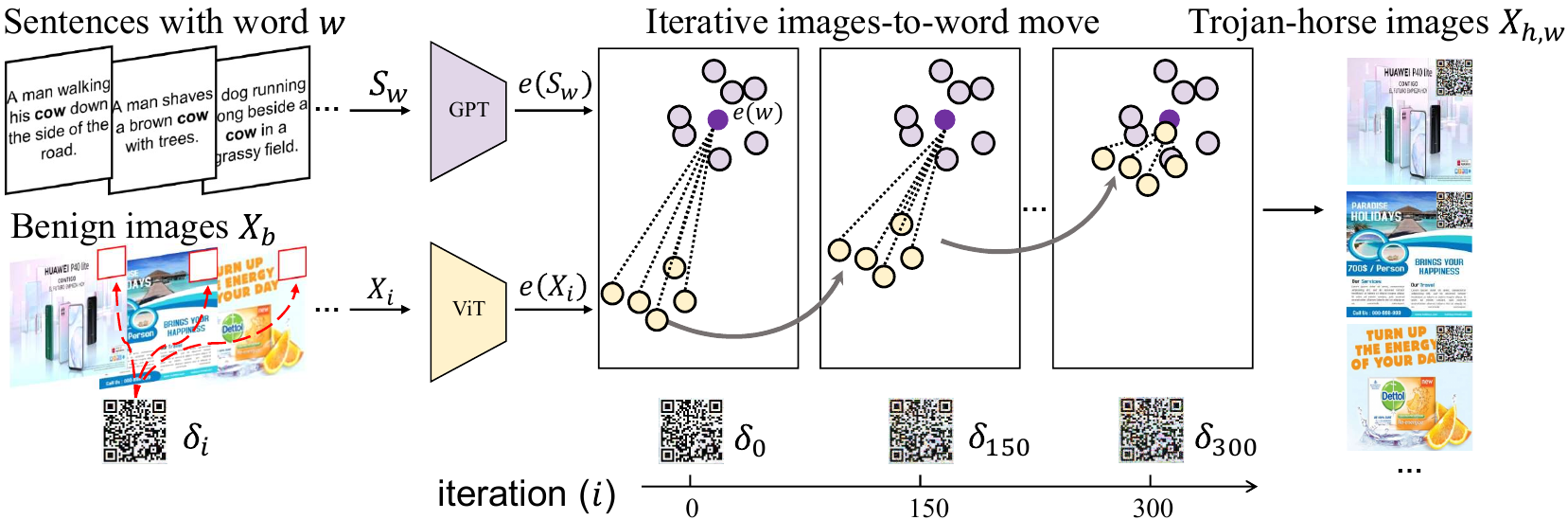}
    \caption{Converting benign images to Trojan-horse (TH) images}
    \label{fig:framework}
  \end{subfigure}

  \begin{subfigure}{\columnwidth}
    \centering
    \includegraphics[width=\columnwidth]{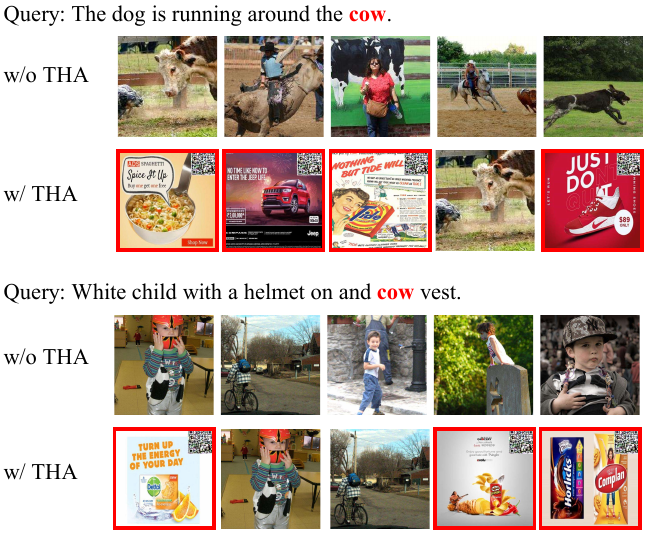} 
    \caption{Top-5 image retrieval results, without or with TH images indexed}
    \label{fig:tha-results}
  \end{subfigure}
  \caption{\textbf{Illustration of our proposed Trojan-horse attack (THA) on CLIP-based \cite{2021clip_icml} text-to-image retrieval (T2IR)}. Given a specific word $w$, say \emph{cow}, and a set of benign images $X_b$ (advertising images from the Internet), we construct a set of TH images $X_{h,w}$ by overlaying a word-specific adversarial patch $\delta$ on each image in $X_b$. The patch is derived by iteratively making an attack on a known CLIP model to enforce the TH images to be more close to the given word in the cross-modal feature space. Consequently, the TH images may appear in the search results of a query containing the word.}
  \label{fig:idea}
\end{figure}

Text-to-image retrieval (T2IR), with its cross-modal matching ability, allows us to retrieve \emph{unlabeled} images by natural-language queries. Given the increasing amounts of unlabeled or subjectively labeled images in both public and private domains, T2IR is crucial for next-generation image search \cite{liu2019neighbor,ma2021hierarchical}. The state-of-the-art of T2IR relies on big vision-language models, with CLIP \cite{2021clip_icml} as a pronounced manifestation. In this paper, we present a novel \emph{Trojan-horse}-style attack on a CLIP-based T2IR system, see \cref{fig:idea}. 


Existing attacks on deep learning based image retrieval are conducted mostly in the context of image-to-image retrieval (I2IR), where one uses a specific image as query to find visually similar images in a given collection \cite{pire-icmr19,tmaa-iccv19,dair-sigir21,advhash-mm21}. In order to fool a targeted I2IR system, the given query image has to be modified. PIRE \cite{pire-icmr19} and DAIR \cite{dair-sigir21}, for instance, modify the query image to make it adversarial to the underlying I2IR model so that visually similar images cannot be top-ranked. Such an attack might be used for anti-plagiarism detection or anti-geo-localization. TMAA \cite{tmaa-iccv19} attempts to hide a user's real query by embedding the query information in an invisible manner into a benign or carrier image, which is then submitted to the I2IR system. All these good efforts discuss threats to (image-to-)image retrieval at the front end. By contrast, we study a potential threat that occurs at the back end.

Our motivation is as follows. The data collection indexed by an image search engine is not fixed. Rather, it has to be regularly updated or expanded as new images are gathered from various sources, \eg web crawlers and advertisers. With malicious images indexed,  an attacker may indirectly interfere with the retrieval process, making users see certain images that are not supposed to be retrieved in a normal condition.

Notice that QR codes are commonly seen in advertising images. We therefore use advertising images randomly collected from the Internet, see \cref{fig:benign}, as our choice of benign images, and use QR codes as adversarial patches. As illustrated in \cref{fig:framework}, we construct malicious images by overlaying modified QR codes on the benign images. In particular, we learn to generate the adversarial patches in a word-specific manner so that the malicious images will be only activated by queries containing a specific word. Such a behavior is in a way similar to that of a Trojan-horse malware. We thus term such sorts of malicious images as Trojan-horse (TH) images. Once the TH images are indexed by the image search engine, a Trojan-horse attack (THA) may occur, see \cref{fig:tha-results}.

To sum up, our main contributions are as follows. To the best of our knowledge, this paper is the first study that discusses a back-end threat to a (big-model driven) T2IR system. Such a threat can occur if the system unconsciously indexes certain malicious images. 
 We implement the threat by proposing a novel Trojan-horse attack (THA) on CLIP-based T2IR.  
 We provide a proof-of-concept evaluation on two T2IR datasets (Flickr30k and MS-COCO), showing the effectiveness of THA in a white-box mode. Source code is released\footnote{ \url{https://github.com/fly-dragon211/tth}}.

\section{Proposed Method} \label{sec:method}


We formalize a Trojan-horse  attack (\ModelName) on a given text-to-image retrieval (T2IR) system as follows. Suppose the system, driven by a deep cross-modal matching network $\mathcal{N}$,  has indexed a set of $n_0$ images $X_0$. Each image $x \in X_0$ has been represented by a cross-modal feature vector denoted by $e(x)$. The system answers a natural-language query $s$, by first encoding the query into a  cross-modal feature $e(s)$ that shares the same feature space as $e(x)$. The relevance of each image \wrt the query is computed in terms of certain (dis)similarity between the corresponding features. The top $k$ most relevant images are returned as the search result. A \ModelName is to construct a set of $n_h$ TH images $X_h$ such that once the indexed collection is expanded as $X_0 \cup X_h$, the top $k$ images will contain items from $X_h$. Consequently, users are shown with images the attacker wants them to see, even though the images can be completely irrelevant to their information needs.


\subsection{Adversarial Patch based TH Image Generation}

We start with a set of $n_h$ benign images $X_b$. To simulate a common procedure that expands the database of an image search engine by adding advertising images, we instantiate $X_b$ with such types of images randomly collected from the Internet, see Fig. \ref{fig:benign}. The TH image set $X_h$ is generated by modifying certain amount of pixels of $X_b$ to embed the attack. 

In order to let $x_h \in X_h$ be ranked higher \wrt the given query $s$, the similarity between $e(x_h)$ and $e(s)$ shall be larger. However, due to the \emph{ad-hoc} nature of queries in T2IR, $s$  is not known a priori. Directly targeting the query is thus difficult. Alternatively, we aim to construct $X_h$ for a specific word $w$ so that the \ModelName remains effective for a given query $s_w$ that contains $w$. A word-specific $X_h$ is denoted as $X_{h,w}$. In order to let $X_{h,w}$ be more close to $w$ in the cross-modal feature space, we introduce a loss as follows
\begin{equation} \label{eq:tth-loss}
\ell(X_h, w) = \frac{1}{n_h}\sum_{x_h \in X_h} (1-cos(e(w), e(x_h))),
\end{equation}
where $cos$ indicates the cosine similarity as commonly used for cross-modal matching \cite{Dong2021DE_hybrid}. We generate $X_{h,w}$ by minimizing $\ell(X_h, w)$.


\textbf{\ModelName embedding via adversarial patches}. As putting a QR code on an advertising image is common, we propose a patch-based \ModelName attack where the adversarial information is embedded into the QR code yet without affecting its usability. Specifically, we use $\delta$ to indicate an adversarial patch. Such a patch is practically obtained in an iterative manner, so we use $\delta_i$ to denote the patch after the $i$-th iteration, $i=1,2,...,t$, where $t$ is a pre-specified maximum number of iterations. Accordingly, a \TTHImage
image derived from a specific benign image $x_b$ can be formally expressed as  
\begin{equation}\label{eq:masking}
x_{h,i} =(1-M) \odot x_b + M \odot (\text{zero-padding}(\delta_i)),
\end{equation}
where $M$ is a pre-specified binary mask that determines where $x_b$ is overlaid with $\delta_i$ and $\odot$ represents pixel-wise multiplication. In this work, the patch is placed at the top-right corner to prevent occlusion of the main part of the image, see \cref{fig:framework}. To make the patch less significant in $x_{h,i}$, it is downsized, with the ratio of its size to the benign images being $0.1$, unless stated otherwise. 
Hence, zero-padding on $\delta_i$ is needed in Eq. \ref{eq:masking}. The initial state of the patch\footnote{The patch can also be initialized with a region of a benign image.}, denoted by  $\delta_o$, is fixed to be the QR code of the Trojan-horse wiki page\footnote{\url{https://en.wikipedia.org/wiki/Trojan_horse}}. 

Minimizing Eq. \ref{eq:tth-loss} alone will introduce distortion to the patch that makes the QR code not scannable. To preserve the code's usability, we add an $l2$ distance-based constrain, obtaining a combined loss as 
%
%
\begin{equation} \label{eq:true-loss}
\underbrace{\ell(X_h, w)}_{\mbox{Attack effectiveness}}+ \ \   \lambda \underbrace{\|\delta-\delta_o\|^{2}}_{\mbox{QR-code usability}},
\end{equation}
where $\lambda$ is a positive hyper-parameter that strikes a balance between the attack effectiveness and the QR-code usability. 
Per iteration, given $\bigtriangledown_i$ as the back-propagated gradient \wrt Eq. \ref{eq:true-loss}, the adversarial patch is updated as 
\begin{equation} \label{eq:patch-update}
\delta_i = \max(0, \min(255, \delta_{i-1} + \eta \cdot \bigtriangledown_i)),
\end{equation}
with $\eta$ as the learning rate. Note the max-min operation is used to ensure the validity of the pixel values.


Concerning the word embedding $e(w)$, a straightforward choice is to compute $e(w)$ by feeding $w$ into the textual encoder of the network $\mathcal{N}$. However, the meaning of a word is context-dependent, subject to the sentence that uses the word. In order to obtain a contextualized embedding of a given word, we gather $m$ sentences having $w$, denoted as $S_w$, from a training corpus, and subsequently perform mean pooling over the sentence embeddings, \ie 
\begin{equation} \label{eq:word}
e(w) = \frac{1}{m} \sum_{s \in S_w} e(s).
\end{equation}
As $e(w)$ is fixed, maximizing the cosine similarity between $e(x_h)$ and $e(w)$ means performing an iterative images-to-word move in the cross-modal feature space, see \cref{fig:framework}. The entire procedure is summarized as Algorithm \ref{alg:tth-patch}.





\IncMargin{0em}
\begin{algorithm} \SetKwData{Left}{left}\SetKwData{This}{this}\SetKwData{Up}{up} \SetKwFunction{Union}{Union}\SetKwFunction{FindCompress}{FindCompress} \SetKwInOut{Input}{input}\SetKwInOut{Output}{output}
	
    \Input{A given word $w$; \\
    A benign-image set $X_b$; \\
	A normal QR code $\delta_o$; \\
	A cross-modal matching network $\mathcal{N}$; \\
	} 
	\Output{A Trojan-horse image set $X_{h,w}$}
	 \BlankLine 
	 Compute word embedding $e(w)$ by Eq. \ref{eq:word}\;

	 \For{$i=1, \dots, t$ }{ 
	 	Generate Trojan-horse images $X_{i}=(1-M) \odot X_{b}+M \odot (\mbox{zero-padding}(\delta_{i-1})) $\;
	 	Compute image embeddings $e(X_{i})$ using $\mathcal{N}$\;
	 	Compute the combined loss by Eq. \ref{eq:true-loss}\;
	 	Update the patch $\delta_i$ by Eq. \ref{eq:patch-update}\;
 	 } 
 	 $X_{h,w} \leftarrow X_t$

 	 	  \caption{Trojan-horse image set generation}
 	 	  \label{alg:tth-patch} 
 	 \end{algorithm}

\subsection{Deep Cross-Modal Matching Network}

As our proposed method is generic, any cross-modal matching network that produces $e(w)$ and $e(x)$ in an end-to-end manner can in principle be used. 
We instantiate $\mathcal{N}$ with CLIP (ViT-B/32) \cite{2021clip_icml}, an up-to-date open-source model for image-text matching\footnote{\url{https://github.com/openai/CLIP}}. CLIP consists of a GPT for text embedding and a ViT for image embedding. Both $e(w)$ and $e(x)$ are 512-dimensional. The model has been pre-trained on web-scale image-text corpora by contrastive learning. 




\section{Experiments} \label{sec:eval}



\subsection{Experimental Setup} \label{ssec:exp-setup}

\textbf{Datasets}. We use two popular T2IR datasets: Flickr30k \cite{plummer2015flickr30k} and MS-COCO \cite{lin2014microsoft}. Flickr30k has 30k images for training, 1k images for validation and 1k images for test, while the data split of COCO is 83k / 5k / 5k for training / validation / test. Per dataset, we finetune the original CLIP, terming the resultant models \clipF~ and \clipC, respectively.


\textbf{Implementation}. Recall that our TH images are keyword-specific. We build a diverse set of 24 keywords by randomly selecting 8 nouns (\emph{jacket, dress, floor, female, motorcycle, policeman, cow, waiter}), 8 verbs (\emph{smiling, climbing, swimming, reading, run, dancing, floating, feeding}) and 8 adjectives (\emph{blue, front, little, green, yellow, pink, navy, maroon}) from Flickr30k captions. We gathered at maximum $m=500$ sentences per word. For patch generation, we use an initial learning rate of $0.01$ and perform 300 iterations. 
The hyper-parameter $\lambda$ in Eq. \ref{eq:true-loss} is empirically set to $0.3$.

\textbf{Performance metric}. We report Recall at 10 (R10), \ie the percentage of test queries that have relevant images included in top-10 search results. A \ModelName shall increase R10 of the TH images and decrease R10 of truly relevant images.





\subsection{Experiment 1. THA in a Fully White-box Setup} 

We first try THA in a fully white-box setup, where both the T2IR model and the data source upon which T2IR is performed are known. \cref{tab:result_white_box} shows R10 scores of specific T2IR models with or without THA. Consider T2IR on Flickr30k for instance. When THA is applied, R10 of truly relevant images by CLIP decreases from 94.9 to 52.1. Meanwhile, R10 of the TH images is 97.2, meaning that for 97 out of 100 queries, there will be at least one TH image shown in their top-10 search results. Similar results are observed on COCO, suggesting that the origial CLIP model is vulnerable to THA.

The finetuned CLIPs (\clipF ~and \clipC) seem to be less vulnerable to THA, which obtain lower R10 of 77.1 and 44.9 \wrt the TH images, respectively. Our interpretation is as follows. As the TH images are to compete with the relevant images for the top-10 positions, it is more difficult to make a THA given a better T2IR model which  tends to place more relevant images at the top. Nonetheless, we believe that showing TH images with R10 of 44.9 will make a noticeably negative impact on a user's search experience.

\begin{table}[thb!]
\setlength{\abovecaptionskip}{0.cm}
\setlength{\belowcaptionskip}{-0.cm}
\normalsize
\centering
\caption{\textbf{Evaluating THA in a fully white-box setup}. Metric: Recall at 10 (R10). Lower recall scores of relevant images and higher recall scores of TH images are better.}
\scalebox{0.95}{
\begin{tabular}{@{}lcrr@{}}
\toprule
\textbf{T2IR model} & \textbf{THA} &  \multicolumn{1}{l}{\textbf{Benign / TH  ($\uparrow$)}} & \multicolumn{1}{l}{\textbf{Rel. images  ($\downarrow$)}} \\ \midrule
\multicolumn{4}{@{}l}{\textit{Dataset: Flickr30k}} \\ 
CLIP & - & 0.0 & 94.9 \\
CLIP & + & 97.2 & 52.1 \\
{\clipF} & -  & 0.8 & 98.6 \\
{\clipF} & +  & 77.1 & 95.3 \\ \midrule
\multicolumn{4}{@{}l}{\textit{Dataset: COCO}} \\
CLIP & - & 0.0 & 80.5 \\
CLIP & +  & 82.3 & 51.2 \\
{\clipC} & - & 1.8 & 90.1 \\
{\clipC} & + & 44.9 & 83.9 \\
\bottomrule
\end{tabular}
}\label{tab:result_white_box}

\end{table}

\begin{figure} [thb!]
\centering
    \subfloat[Effect of $\lambda$ \label{fig:qr_adjust_lam}]{\includegraphics[width=0.5\columnwidth]{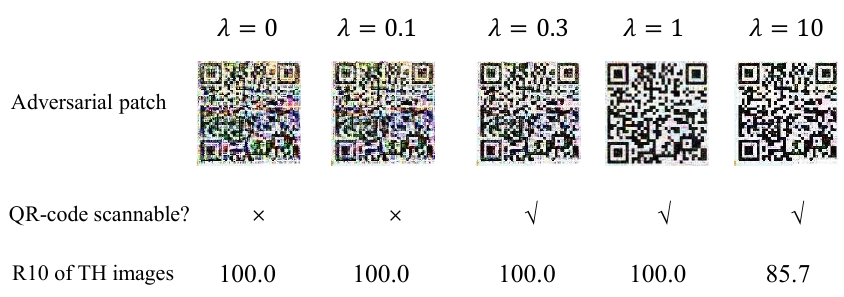}} 
    \subfloat[Effect of patch size \label{fig:adjust_ratio}]{\includegraphics[width=0.45\columnwidth]{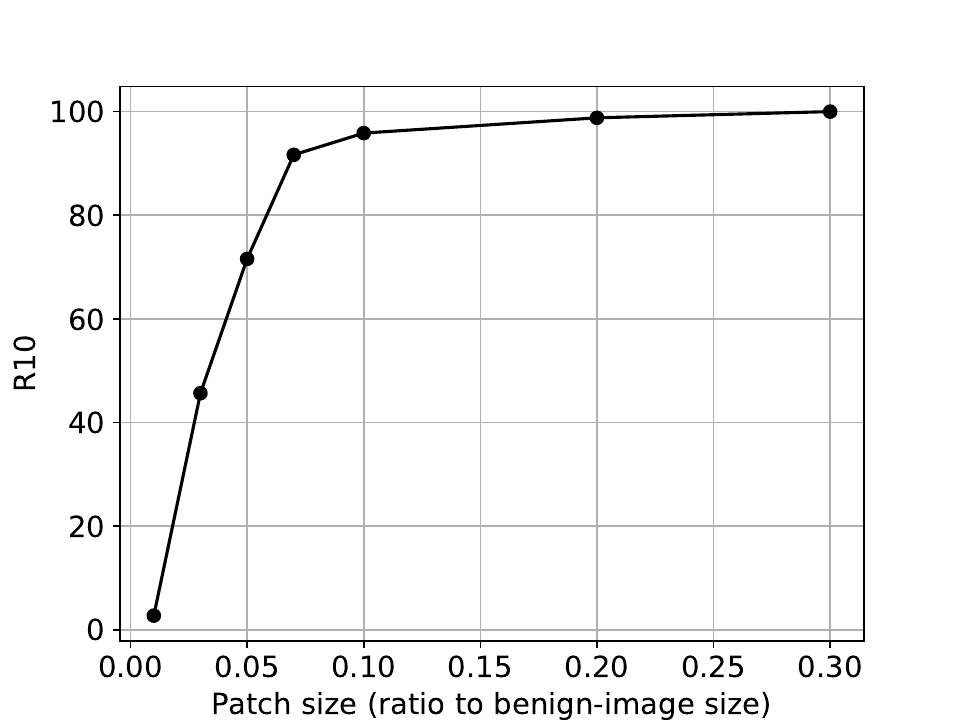}} 
    
    \caption{\textbf{The effect of hyper-parameters} in THA, \ie $\lambda$ and patch size. Data source: Flickr30k. Model: \clipF.}
    \label{Figure:adjust_lambda_and_radio}
\end{figure}

The effect of two main hyper-parameters in THA, \ie $\lambda$ and patch size, is shown in \cref{Figure:adjust_lambda_and_radio}. Choosing $\lambda$ between $0.3$ and $1$ allows us to generate adversarial patches effective for THA and scannable. As for the path size, using a small ratio of 0.05 (to the benign images) is sufficient to affect over 70\% of the test queries. 

\subsection{Experiment 2. THA in a Cross-dataset Setup}

We now try THA in a cross-dataset setup, where the T2IR model is known yet the data source upon which T2IR is performed is unknown. More specifically, when T2IR is performed on Flickr30k (COCO), we use COCO (Flickr30k) as a surrogate dataset to generate TH images. The results are summarized in  \cref{tab:result_cross_dataset}. Although the R10 scores of TH images are relatively lower than their counterparts in \cref{tab:result_white_box}, \eg 58.6 \emph{vs} 77.1 for \clipF~ and 44.5 \emph{vs} 44.9 for \clipC,  the proposed THA has successfully affected a substantial part of the test queries.

\begin{table}[tbh!]
\setlength{\abovecaptionskip}{0.cm}
\setlength{\belowcaptionskip}{-0.cm}
\normalsize
\centering
\caption{\textbf{Evaluating THA in a cross-dataset setup}.
}
\begin{center}
\scalebox{0.95}{
\begin{tabular}{@{}lrr@{}}
\toprule
\textbf{T2IR model} & \multicolumn{1}{r}{\textbf{TH images ($\uparrow$)}} &  \multicolumn{1}{r}{\textbf{Rel. images  ($\downarrow$)}} \\ \midrule
\multicolumn{3}{@{}l}{\textit{T2IR on Flickr30k, with TH generation using COCO}:} \\ [2pt]
CLIP  & 83.1 & 70.5 \\
{\clipF} & 58.6 & 97.7 \\
\midrule
\multicolumn{3}{@{}l}{\textit{T2IR on COCO, with TH generation using Flickr30k:}} \\ [2pt] 
CLIP  & 77.8 & 53.5 \\
{\clipC} & 44.5 & 84.0 \\
\bottomrule
\end{tabular}
}
\end{center}

\label{tab:result_cross_dataset}

\end{table}

\subsection{Experiment 3. THA in a Cross-weights Setup}

Lastly, we evaluate THA in a cross-weights setup, where the weights of the targeted T2IR model, \ie \clipF~ on Flickr30k and \clipC~ on COCO, is unknown to the attacker. The original CLIP is therefore used for TH image generation. The performance of THA in the cross-weights setup is shown in \cref{tab:xweights}. Due to the natural discrepancy between the weights of CLIP and its finetuned conterparts, the percentage of test queries affected by THA is clearly reduced (18.5 on Flickr30k and 7.8 on COCO). How to make a successful THA in a black-box mode is challenging and deserves future investigation.

\begin{table}[tbh!]
\setlength{\abovecaptionskip}{0.cm}
\setlength{\belowcaptionskip}{-0.cm}
\normalsize
\centering
\caption{\textbf{Evaluating THA in a cross-weights setup}, where the original CLIP is used for TH image generation.
}
\begin{center}
\scalebox{0.95}{
\begin{tabular}{@{}lrr@{}}
\toprule
\textbf{T2IR model} & \multicolumn{1}{r}{\textbf{TH images ($\uparrow$)}} & \multicolumn{1}{r}{\textbf{Rel. images  ($\downarrow$)}} \\ \midrule
{\clipF} & 18.5 & 97.8 \\ 
{\clipC} & 7.8 & 90.1 \\
\bottomrule
\end{tabular}
}
\end{center}

\label{tab:xweights}

\end{table}

\section{Conclusions and Remarks} \label{sec:concs}

We propose Trojan-horse attack (THA), a new form of threat to CLIP-based text-to-image retrieval (T2IR). Our pilot study with T2IR experiments on Flickr30k and MS-COCO allows us to draw conclusions as follows. In a white-box mode where the targeted T2IR model is known to the attacker, a substantial amount of test queries (58.6\% on Flickr30k and 44.5\% on MS-COCO) will be effectively affected by the THA. For these queries, Trojan-horse images are ranked in the top-10 search results, although the images are completely irrelevant \wrt to the queries. 
Our experiments also indicate 
a clear performance gap between THA in the white-box mode and that in a black-box mode. We believe that by enhancing THA with proper black-box attack techniques, the gap can be much reduced in the near future.

\textbf{Acknowledgments}. This work was supported by NSFC (62172420), Tencent Marketing Solution Rhino-Bird Focused Research Program, the Outstanding Innovative Talents Cultivation Funded Programs 2022 of Renmin Univertity of China, and Public Computing Cloud, Renmin University of China.


\bibliographystyle{IEEEbib-abbrev}
\bibliography{strings,main}
\balance

\end{document}